\documentclass[10pt]{llncs}

\usepackage[small,compact]{titlesec}
\usepackage{epsfig}
\usepackage[utf8]{inputenc}
\usepackage{algorithm}
\usepackage{algorithmic}
\usepackage{graphicx}
\usepackage{amsmath,amsfonts,amssymb}
\usepackage{latexsym}
\usepackage{color}
\usepackage{mathrsfs}
\usepackage{subfigure}
\usepackage{slashbox}
\usepackage{multirow}
\usepackage{wrapfig}
\usepackage[table]{xcolor}
\usepackage[square, numbers]{natbib}
\usepackage{cancel}
\usepackage{bbm}

\makeatletter
\renewcommand\bibsection
{
	\section*{\refname
	\@mkboth{\MakeUppercase{\refname}}{\MakeUppercase{\refname}}}
}
\makeatother

\newcommand{\beq}{\begin{equation}}
\newcommand{\eeq}{\end{equation}}

\newcommand{\bea}{\begin{eqnarray}}
\newcommand{\eea}{\end{eqnarray}}
\newcommand{\bean}{\begin{eqnarray*}}
\newcommand{\eean}{\end{eqnarray*}}

\newcommand{\bit}{\begin{itemize}}
\newcommand{\eit}{\end{itemize}}
\newcommand{\ben}{\begin{enumerate}}
\newcommand{\een}{\end{enumerate}}

\newcommand{\argmax}{\arg\,\max}
\newcommand{\argmin}{\arg\,\min}

\usepackage{amssymb}
\newenvironment{prf}{\noindent{\bf Proof:} \hspace*{1mm}}{ \hspace*{\fill} $\Box$ }

\begin{document}




\title{Mechanism Design for Time Critical and Cost Critical Task Execution via Crowdsourcing}

\author{Swaprava Nath\inst{1} \and Pankaj Dayama\inst{2} \and Dinesh Garg\inst{3} \and  \\ Y. Narahari\inst{1} \and James Zou\inst{4}}
\institute{Department of Computer Science and Automation, Indian Institute of Science, Bangalore, India, \texttt{\{swaprava, hari\}@csa.iisc.ernet.in}
 \and IBM India Research Lab, Bangalore, India, \texttt{pankajdayama@gmail.com}
\and IBM India Research Lab, New Delhi, India, \texttt{garg.dinesh@in.ibm.com}
\and Harvard School of Engineering and Applied Sciences, Cambridge, MA, \texttt{jzou@fas.harvard.edu}
}

\maketitle
\begin{abstract}
An exciting application of crowdsourcing is to use social networks in complex task execution. In this paper, we address the problem of a planner who needs to incentivize agents within a network in order to seek their help in executing an {\em atomic task} as well as in recruiting other agents to execute the task. We study this mechanism design problem under two natural resource optimization settings: (1) cost critical tasks, where the planner's goal is to minimize the total cost, and (2) time critical tasks, where the goal is to minimize the total time elapsed before the task is executed. We identify a set of desirable properties that should ideally be satisfied by a crowdsourcing mechanism. In particular, {\em sybil-proofness} and {\em collapse-proofness} are two complementary properties in our desiderata. We prove that no mechanism can satisfy all the desirable properties simultaneously. This leads us naturally to explore approximate versions of the critical properties. We focus our attention on approximate sybil-proofness and our exploration leads to a parametrized family of payment mechanisms which satisfy collapse-proofness. We characterize the approximate versions of the desirable properties in cost critical and time critical domain.
\end{abstract}
\section{Introduction} \label{sec:intro}
Advances in the Internet and communication technologies have made it possible to harness the wisdom and efforts from a sizable portion of the society towards accomplishing tasks which are otherwise herculean. Examples include labeling millions of images, prediction of stock markets, seeking answers to specific queries, searching for objects across a wide geographical area, etc. This phenomenon is popularly known as {\em crowdsourcing} (for details, see \citet{Surowiecki} and \citet{Howe}). {\em Amazon Mechanical Turk} is one of the early examples of online crowdsourcing platform. The other example of such online crowdsourcing platforms include {\em oDesk, Rent-A-Coder, kaggle, Galaxy Zoo,} and {\em Stardust@home}.

In recent times, an explosive growth in online social media has given a novel twist to crowdsourcing applications where participants can exploit the {\em underlying social network} for inviting their friends to help executing the task.
In such a scenario, the task owner initially recruits individuals from her immediate network to participate in executing the task. These individuals, apart from attempting to execute the task by themselves, recruit other individuals in their respective social networks to also attempt the task and further grow the network. An example of such applications include the DARPA {\em Red Balloon Challenge}~\citep{DARPA}, {\em DARPA CLIQR quest} \citep{DARPA-cliqr}, {\em query incentive networks} \citep{Kleinberg05}, and {\em multi-level marketing} \citep{Emek11}. The success of such crowdsourcing applications depends on providing appropriate incentives to individuals for both (1) executing the task by themselves and/or (2) recruiting other individuals. Designing a proper incentive scheme ({\em crowdsourcing mechanism}) is crucial to the success of any such crowdsourcing based application. In the red balloon challenge, the winning team from MIT successfully demonstrated that a crowdsourcing mechanism can be employed to accomplish such a challenging task (see \cite{pickard-etal11MIT}).

A major challenge in deploying such crowdsourcing mechanisms in realistic settings is their vulnerability to different kinds of manipulations (e.g.\ false name attacks, also known as {\em sybil attacks\/} in the literature) that rational and intelligent participants would invariably attempt. This challenge needs to be addressed in a specific manner for a specific application setting at the time of designing the mechanism. The application setting is characterized, primarily, by the nature of the underlying task and secondly, by the high level objectives of the designer. Depending on the nature of the underlying task, we can classify them as follows.

 \noindent {\bf Viral Task.} A viral task is the one where the designer's goal is to involve as many members as possible in the social network. This kind of tasks do not have a well defined stopping criterion. Examples of such a task include viral marketing, multi-level marketing, users of a social network participating in an election, etc. 

 \noindent {\bf Atomic Task.} An atomic task is one in which occurrence of a particular event (typically carried out by a single individual) signifies the end of the task. By definition, it comes with a well defined measure of success or accomplishment. Examples of an atomic task include the DARPA Red Balloon Challenge, DARPA CLIQR quest, query incentive networks, and transaction authentication in Bitcoin system~\citep{Babaioff2012}. 

In this paper, we focus on the problem of designing crowdsourcing mechanisms for atomic tasks such that the mechanisms are robust to any kind of manipulations and additionally achieve the stated objectives of the designer.

\section{Prior Work} \label{sec:related}
Prior work can be broadly classified into two categories based on the nature of the underlying task - viral or atomic. \\
{\bf Viral Task:} The literature in this category focuses, predominantly, on the problem of multi-level marketing. \citet{Emek11} and \citet{Drucker2012} have analyzed somewhat similar models for multi-level marketing over a social network. In their model, the planner incentivizes agents to promote a product among their friends in order to increase the sales revenue. While \cite{Emek11} shows that the geometric reward mechanism uniquely satisfies many desirable properties except false-name-proofness, \cite{Drucker2012} presents a {\em capping reward mechanism} that is locally sybil-proof and collusion-proof. The collusion here only considers creating fake nodes in a collaborative way. In all multi-level marketing mechanisms, the revenue is generated {\em endogenously} by the participating nodes, and a fraction of the revenue is redistributed over the referrers. On slightly different kind of tasks, \citet{Conitzer10} proposes mechanisms that are robust to false-name manipulation for applications such as facebook inviting its users to vote on its future terms of use. Further, \citet{Yu06} proposes a protocol to limit corruptive influence of sybil attacks in P2P networks by exploiting insights from social networks.\\
{\bf Atomic Task:} The red-balloon challenge \citep{DARPA}, query incentive networks \citep{Kleinberg05}, and transaction authentication in Bitcoin system \citep{Babaioff2012} are examples of {\em atomic tasks}.
The reward in such settings is {\em exogenous}, and hence the strategic problems are different from the viral tasks such as multi-level marketing. Sybil attacks still pose a problem here. \citet{pickard-etal11MIT} proposed a novel solution method for Red Balloon challenge and can be considered as an early work that motivated the study of strategic aspects in crowdsourcing applications. \cite{Babaioff2012} provides an {\em almost uniform} mechanism where sybil-proofness is guaranteed via iterated elimination of weakly dominated strategies.
The work by \citet{Kleinberg05} deals with a branching process based model for query incentive networks
and proposes a decentralized reward mechanism for the nodes along the path from the root to the node who answers the query.
\section{Contributions and Outline} \label{sec:contribution}
In this paper, we propose design of crowdsourcing mechanisms for atomic tasks such that the mechanisms are robust to any kind of manipulations and additionally achieve the stated objectives of the designer.
Our work is distinct from the existing body of related literature in the following aspects.\\
{\bf (1) Collapse-Proofness:} We discover that agents can exhibit an important strategic behavior, namely {\em node collapse attack}, which has not been explored in literature. Though the sybil attack has been studied quite well, a sybil-proof mechanism cannot by itself prevent multiple nodes colluding and reporting as a single node in order to increase their collective reward. A node collapse behavior of the agents is undesirable because, (i) it increases cost to the designer, (ii) the distribution of this additional payment creates a situation of bargaining among the agents, hence is not suitable for risk averse agents, and (iii) it hides the structure of the actual network, which could be useful for other future purposes. A node collapse is a form of collusion, and it can be shown that the sybil-proof mechanisms presented in both \cite{Babaioff2012} and \cite{Drucker2012} are vulnerable to collapse attack. In this paper, in addition to sybil attacks, we also address the problem of {\em collapse attacks} and present mechanisms that are collapse-proof.\\
{\bf (2) Dominant Strategy Implementation:} In practical crowdsourcing scenarios, we cannot expect all the agents to be fully {\em rational and intelligent}. We, therefore, take a complementary design approach, where instead of satisfying various desirable properties (e.g. sybil-proofness, collapse-proofness) in the Nash equilibrium sense,~\footnote{For example, the solution provided by \citet{Babaioff2012} guarantees sybil-proofness only in Nash equilibrium and not in dominant strategies.} we prefer to address a {\em approximate versions} of the same properties, and design {\em dominant strategy} mechanisms. If a mechanism satisfies an approximate version of a cheat-proof property then it means the loss in an agents' utility due to him following a non-cheating behavior is bounded (irrespective of what others are doing).\\
{\bf (3) Resource Optimization Criterion:} The present literature mostly focuses on the design of a crowdsourcing mechanism satisfying a set of desirable cheat-proof properties. The feasible set could be quite large in many scenarios and hence a further level of optimization of the resources would be a natural extension. In this paper, we demonstrate how to fill this gap by analyzing two scenarios - (1) cost critical tasks, and (2) time critical tasks.

A summary of our specific contributions in this paper is as follows.
\begin{enumerate}
\item We identify a set of desirable properties, namely (1) {\em Downstream Sybil-proofness (DSP)}, (2) {\em Collapse-proofness (CP)}, (3)  {\em Strict Contribution Rationality (SCR)}, (4) {\em Budget Balance (BB)}, and (5) {\em Security to Winner (SEC)}.
 \item We first prove that not all properties above (in fact, even subsets of these properties) are simultaneously satisfiable (Theorem \ref{prop:impossible}).  

 \item We next prove  a possibility result which shows that DSP, SCR, CP, and BB can be simultaneously satisfied but under a very restrictive mechanism (Theorem \ref{prop:wintakeall}).
 \item
Next, we propose dominant strategy mechanisms for {\em approximate versions} of these properties, which is complementary to the solution provided by \citet{Babaioff2012} that guarantees sybil-proofness in Nash equilibrium. In particular, we  define the notion of $\epsilon$-DSP, $\delta$-SCR, and $\gamma$-SEC. The need for defining an approximate version of the CP property does not arise since all the proposed mechanisms satisfy exact collapse-proofness.
 \item The approximate versions help expand the space of feasible mechanisms, leading us naturally to the following question: {\em How should the mechanism designer (task owner or planner) choose a particular mechanism from a bunch of possibilities\/}? We ask this question in two natural settings: (a) {\em cost critical tasks}, where the goal is to minimize the total cost, (b) {\em time critical tasks}, where the goal is to minimize the total time for executing the task~\footnote{Note, {\em query incentive networks} \citep{Kleinberg05} and {\em multi-level marketing} \citep{Emek11} fall under the category of cost critical tasks, while search-and-rescue operations such as red balloon challenge \citep{DARPA} fall under that of time critical tasks.}. We provide characterization theorems (Theorems \ref{thm:main-full} and \ref{thm:maxleaf}) in both the settings for the mechanisms satisfying  approximate properties ($\epsilon$-DSP, $\delta$-SCR, and $\gamma$-SEC)  in conjunction with the CP property.
\end{enumerate}
To the best of our knowledge, this is the first attempt at providing approximate sybil-proofness and exact collapse-proofness in dominant strategies with certain additional fairness guarantees ($\delta$-SCR and $\gamma$-SEC).
\section{The Model} \label{sec:model}
Consider a planner (such as DARPA) who needs to get an atomic task executed. The planner recruits a set of agents and asks them to execute the task.
The recruited agents can try executing the task themselves or in turn forward the task to their friends and acquaintances who have not been offered this deal so far, thereby recruiting them into the system. If an agent
receives separate invitations from multiple nodes to join their network, she can accept exactly one invitation. Thus, at any point of time, the recruited agents network is a tree. The planner stops the process as soon as the atomic task gets executed by one of the agents and offers rewards to the agents as per a centralized monetary reward scheme, say $R$. Let $T = (V_T, E_T)$ denote the final recruitment tree when the atomic task gets executed by one of the recruited agents. In $T$, the agent who executes the atomic task first is referred to as the \emph{winner}. Let us denote the winner as $w \in V_T$. The unique path from the winner to the root is referred to as the \emph{winning chain}. We consider the mechanisms where only winning chain receives positive payments.

For our setting, we assume that the planner designs the centralized reward mechanism $R$, which assigns a non-negative reward to every node in the winning chain and zero to all other nodes. Hence, we can denote the reward mechanism as a mapping  $R : \mathbb{N} \times \mathbb{N} \to \mathbb{R}^+$  where $\mathbb{N}$ is the set of natural numbers and $\mathbb{R}^+$ is the set of nonnegative reals. In such a mechanism, $R(k,t), \ k \leq t$ denotes the reward of a node which is at depth $k$ in the winning chain, where length of the winning chain is $t$.  The payment is made only after completion of the task. Note, this reward mechanism is anonymous to node identities and the payment is solely dependent on their position in $T$. Throughout this paper, we would assume that the payment to all nodes of any \emph{non-winning chain} is zero. Hence, all definitions of the desirable properties apply only to the winning chain.

An example of such a reward mechanism is the geometric payment used by~\citet{Emek11} and~\citet{pickard-etal11MIT}. These mechanisms pay the largest amount to the winner node and geometrically decrease the payment over
the path to the root. This class of mechanisms are susceptible to sybil attacks. For example, the winning node can create a long chain of artificial nodes, $\{x_1, ..., x_m\}$, and report that $x_i$ recruits $x_{i+1}$ and $x_m$ is the winner. Then each fake $x_i$ would extract payment from the mechanism.
\subsection{Desirable Properties}
An ideal reward mechanism of our model should satisfy several desirable properties. In what follows, we have listed down a set of very natural properties that must be satisfied by an ideal mechanism under dominant strategy equilibrium.
\begin{definition}[Downstream Sybilproofness, DSP]
\label{def:down-sybil}
 Given the position of a node in a recruitment tree, a reward mechanism $R$ is called \emph{downstream sybil-proof}, if the node cannot gain by adding fake nodes below itself in the current subtree (irrespective of what other are doing). Mathematically,
 \begin{eqnarray}
  & R(k,t) \geq \sum_{i=0}^n R(k+i, t+n) \quad \forall k \leq t, \forall t, n. \label{eq:down}
 \end{eqnarray} 
\end{definition}
\begin{definition}[Budget Balance, BB]
 \label{def:bb}
 Let us assume the maximum budget allocated by the planner for executing an atomic task is $R_{\max}$. Then, a mechanism $R$ is budget balanced if,
 \begin{eqnarray}
 & \sum_{k=1}^t R(k,t) \leq R_{\max}, \quad \forall t. \label{eq:bb}
 \end{eqnarray}
\end{definition}

\begin{definition}[Contribution Rationality, CR]
\label{def:cont-rational}
 This property ensures that a node gets non-negative payoff whenever she belongs to the winning chain.
We distinguish between strict and weak versions of this property as defined below.
For all $t \geq 1$,

\textbf{Strict Contribution Rationality (SCR):}
\begin{equation}
 R(k,t) > 0, \quad \forall k \leq t, \mbox{ if $t$ is the winning chain.} \label{eq:str-rational}
\end{equation}

\textbf{Weak Contribution Rationality (WCR):}
\begin{align}
 & R(k,t) \geq 0, \quad \forall k \leq t-1, \mbox{if $t$ is the winning chain.} \nonumber \\
 & R(t,t) > 0, \qquad \mbox{winner gets positive reward.} \label{eq:weak-rational}
\end{align}
\end{definition}
DSP ensures that an agent in the network cannot gain additional payment by creating fake identities and pretending to have recruited these nodes. SCR ensures that nodes have incentive to recruit, since all members of the winning chain are rewarded.

There are many reward mechanisms that satisfy these three properties. For example, let us consider a mechanism that diminishes the rewards geometrically in both $k$ and $t$, i.e. $R(k,t) = \frac{1}{2^{k+t}} \cdot R_{\max}$. This mechanism pays heavy to the nodes near the root and less near the leaf. We call this class of mechanisms as {\em top-down} mechanisms. This mechanism satisfies DSP, BB, and SCR properties for any finite $t$. However, the best response strategy of the agents in this type of mechanisms could introduce other kinds of undesirable behavior. For example, the agents of any chain would be better off by colluding among themselves and representing  themselves as a single node in front of the the designer, since if the winner emerges from that particular chain, they would gain more collective reward than they could get individually. We call this  {\em node collapse problem}. This introduces a two-fold difficulty. First, the designer cannot learn the structure of the network that executed the task, and hence  cannot use the network structure for future applications. Second, she ends up paying more than what she should have paid for a true network. Hence, in the scenario where designer is also willing to minimize the expenditure, she would like to have \emph{collapse-proofness}.

\begin{definition}[Collapse-Proofness, CP]
 Given a depth $k$ in a winning chain, a reward mechanism $R$ is called \emph{collapse-proof}, if the subchain of length $p$ down below $k$ collectively cannot gain by collapsing to depth $k$ (irrespective of what others are doing). Mathematically,
 \begin{eqnarray}
  & \sum_{i=0}^{p} R(k+i, t) \geq R(k,t-p) \quad \forall k +p \leq t, \forall t. \label{eq:collapse}
 \end{eqnarray}
\end{definition}
In the following section, we will show that some of these properties are impossible to satisfy
together. To this end, we need to define a class of mechanisms, called {\em Winner Takes All (WTA)},
where the winning node receives a positive reward and all other nodes get zero reward.
\begin{definition}[WTA Mechanism]
 A reward mechanism $R$ is called  WTA mechanism if $R_{\max} \geq R(t,t) > 0$, and $R(k, t) = 0, \ \forall k < t.$
\end{definition}
\section{Impossibility and Possibility Results} \label{sec:results}
\begin{theorem}[Impossibility Result] \label{prop:impossible}
 No reward mechanism can satisfy DSP, SCR, and CP together.
\end{theorem}
\begin{prf}
 Suppose the reward mechanism $R$ satisfies DSP, SCR, and CP. Then by CP, let us put $t \leftarrow t+n$ and $p \leftarrow n$ in Equation~\ref{eq:collapse}, and we get, $\sum_{i=0}^{n} R(k+i, t+n) \geq R(k,t+n-n) = R(k,t), \ \forall k \leq t, \forall t, n.$ This is same as Equation~\ref{eq:down} with the inequality reversed. So, to satisfy DSP and CP together, the inequalities reduce to the following equality.
\begin{eqnarray}
 & R(k,t) = \sum_{i=0}^{n} R(k+i, t+n), \forall k \leq t, \forall t, n. \label{eq:equality}
\end{eqnarray}
Now we use the following substitutions, leading to the corresponding equalities.
\begin{align}
 & \mbox{put } k \leftarrow t-2, t \leftarrow t-2, n \leftarrow 2, \mbox{ to get}, \nonumber \\
& R(t-2,t-2) = R(t-2, t) + R(t-1,t) + R(t,t) \label{eq:1} \\
 &\mbox{put } k \leftarrow t-1, t \leftarrow t-1, n \leftarrow 1, \mbox{ to get}, \nonumber \\
& R(t-1,t-1) = R(t-1,t) + R(t,t) \label{eq:2} \\
& \mbox{put } k \leftarrow t-2, t \leftarrow t-2, n \leftarrow 1, \mbox{ to get}, \nonumber \\
& R(t-2, t-2) = R(t-2, t-1) + R(t-1,t-1) \label{eq:3} \\
& \mbox{put } k \leftarrow t-2, t \leftarrow t-1, n \leftarrow 1, \mbox{ to get}, \nonumber \\
& R(t-2, t-1) = R(t-2, t) + R(t-1,t) \label{eq:4}
\end{align}
Substituting the value of Eq.~\ref{eq:2} on the RHS of Eq.~\ref{eq:3},
\begin{equation}
 R(t-2,t-2) = R(t-2, t-1) + R(t-1,t) + R(t,t) \label{eq:5}
\end{equation}
Substituting Eq.~\ref{eq:5} on the LHS of Eq.~\ref{eq:1} yields
\begin{equation}
 R(t-2,t)=R(t-2, t-1) \label{eq:consol}
\end{equation}
From Eq.~\ref{eq:consol} and Eq.~\ref{eq:4}, we see that,
\begin{equation}
 R(t-1,t) =0. \label{eq:zero}
\end{equation}
which contradicts SCR.
\end{prf}\\
From the above theorem and the fact that additional properties reduce the space
of feasible mechanisms, we obtain the following corollary.
\begin{corollary}
 It is impossible to satisfy DSP, SCR, CP, and BB together.
\end{corollary}
\begin{theorem}[Possibility Result] \label{prop:wintakeall}
 A mechanism satisfies DSP, WCR, CP and BB iff it is a WTA mechanism.
\end{theorem}
\begin{prf}
 ($\Leftarrow$) It is easy to see that WTA mechanism satisfies DSP, WCR, CP and BB. Hence, it suffices to investigate
the other direction.

 ($\Rightarrow$) From Equations~\ref{eq:2} and \ref{eq:zero}, we see that, $R(t-1,t-1) = R(t,t),$ which is true for any $t$. By induction on the analysis of Theorem~\ref{prop:impossible} for length $t-1$ in place of $t$, we can show that $R(t-2,t-1) = 0.$ But, by Eq.~\ref{eq:consol}, $R(t-2,t-1) = R(t-2,t)$. Hence, $R(t-2,t) = 0$. Inductively, for all $t$ and for all $k < t$, $R(k,t) = 0$. It shows that for all non-winner nodes, the reward would be zero. So, we can assign any positive reward to the winner node and zero to all others, which is precisely the WTA mechanism. This proves that for WCR, the reward mechanism that satisfies DSP, CP and BB must be a WTA mechanism.
\end{prf}
\section{Approximate Versions of Desirable Properties} \label{sec:desirable}
The results in the previous section are disappointing in that the space of mechanisms satisfying desirable properties is extremely restricted (WTA being the only one). This suggests two possible ways out of this situation. The first route is to compromise on stronger equilibrium notion of dominant strategy and settle for a slightly weaker notion such as Nash equilibrium. The other route could be to weaken these stringent properties related to cheat-proofness and still look for a dominant strategy equilibrium. We choose to go by the later way because Nash equilibrium makes assumptions of all players being rational and intelligent which may not be true in crowdsourcing applications. Therefore, we relax some of the desirable properties to derive their approximate versions. We begin with approximation of the DSP property.
\begin{definition}[$\epsilon$ - Downstream Sybilproofness, $\epsilon$-DSP]
\label{def:epsilon-DSP}
 Given the position of the node in a tree, a payment mechanism $R$ is called \emph{$\epsilon$ - DSP}, if the node cannot gain by more than a factor of $(1+\epsilon)$ by adding fake nodes below herself in the current subtree (irrespective of what others are doing). Mathematically,
 \begin{equation}
  (1+\epsilon) \cdot R(k,t) \geq \mbox{$\sum_{i=0}^n R(k+i, t+n), \forall k \leq t, \forall t, n.$} \label{eq:e-dsp}
 \end{equation}
\end{definition}
\begin{theorem}
 \label{thm:existence}
 For all $\epsilon > 0$, there exists a mechanism that is $\epsilon$-DSP, CP, BB, and SCR.
\end{theorem}
\begin{prf}
 The proof is constructive. Let us consider the following mechanism: set $R(t,t) = (1-\delta) \cdot R_{\max}, \forall \ t$, the reward to the winner, where $\delta \leq \frac{\epsilon}{1+\epsilon}$. Also, let $R(k,t) = \delta \cdot R(k+1,t) = \delta^{t-k} \cdot R(t,t) = \delta^{t-k} (1-\delta) R_{\max}, k \leq t-1$. By construction, this mechanism satisfies BB. It is also SCR, since $\delta \in (0,1)$. It remains to show that this satisfies $\epsilon$-DSP and CP. Let us consider,
 \begin{align*}
 \lefteqn{\mbox{$\sum_{i=0}^n R(k+i,t+n) = \sum_{i=0}^n \delta^{t+n-k-i} \cdot R(t+n,t+n)$}} \\
      &= \delta^{t-k} \cdot (1+\delta + \dots + \delta^n) \cdot (1-\delta) R_{\max} \\
      &= R(k,t) \cdot (1+\delta + \dots + \delta^n) \\
      &\leq R(k,t) \cdot \frac{1}{1-\delta} \leq (1+\epsilon) \cdot R(k,t), \quad \mbox{since } \delta \leq \frac{\epsilon}{1+\epsilon}.
 \end{align*}
This shows that this mechanism is $\epsilon$-DSP. Also,
 \begin{align*}
  \lefteqn{\mbox{$\sum_{i=0}^p R(k+i,t) = \sum_{i=0}^p \delta^{t-k-i} \cdot R(t,t)$}} \\
      &= \mbox{$\sum_{i=0}^{p-1} \delta^{t-k-i} \cdot R(t,t)$} + \underbrace{\delta^{t-k-p} \cdot R(t,t)}_{ = R(k,t-p)} \geq R(k,t-p)
 \end{align*}
This shows that this mechanism is CP as well.
\end{prf}\\\\
\textbf{Discussion:}
\vspace*{-0.2cm}
\begin{itemize}
\item Above theorem suggests that merely weakening the DSP property allows a way out of the impossibility result given in Theorem \ref{prop:impossible}. One can try weakening the CP property analogously (instead of DSP) and check for the possibility/impossibility results. This we leave as an interesting future work.
\item One may argue that no matter how small is $\epsilon$, as long we satisfy $\epsilon$-DSP property, an agent would always find it beneficial to add as many sybil nodes as possible. However, in real crowdsourcing networks, there would be a non-zero cost involved in creating fake nodes and hence there must be a tipping point so that agent's net gain would increase till he creates that many sybil nodes but starts declining after that. Note, it is impossible for an agent to compute the tipping point a priori as his own reward is uncertain at the time of him getting freshly recruited by someone and he trying to create sybil nodes. Therefore, in the face of this uncertainty, the agent can assure himself of a bounded regret if he decides not to create any sybil nodes.
\end{itemize}
\subsection{Motivation for $\delta$-SCR and $\gamma$-SEC}
\label{sec:fair}
As per previous theorem, the class of mechanisms that satisfy $\epsilon$-DSP, CP, BB, and SCR is quite rich. However, the exemplar mechanism of this class, which was used in the proof of this theorem, prompts us to think of the following  undesirable consequence -
the planner can assign arbitrarily low reward to the winner node and still manage to satisfy all these properties. This could discourage the agents from putting in  effort by themselves for executing the task. Motivated by this considerations, we further extend the SCR property by relaxing it to $\delta$-SCR and also introduce an additional property, namely {\em Winner's $\gamma$ Security ($\gamma$-SEC)}.
\begin{definition}[$\delta$ - Strict Contribution Rationality, $\delta$-SCR]
\label{def:delta-cont-rational}
This ensures that a node in the winning chain gets at least $\delta \in (0,1)$ fraction of her successor. Also the the winner gets a positive reward. For all $t \geq 1$,
\begin{align}
 &  R(k,t) \geq \delta R(k+1,t), \forall k \leq t-1, \mbox{ $t$: winning chain.} \nonumber \\
 &  R(t,t) > 0, \qquad \quad \mbox{ winner gets positive reward.} & \label{eq:delta-scr}
\end{align}
\end{definition}
\begin{definition}[Winner's $\gamma$ Security, $\gamma$-SEC]
This ensures that payoff to the winning node is at least $\gamma$ fraction of the total available budget.
\begin{equation}
 R(t,t) \geq \gamma \cdot R_{\max}, \quad \mbox{ $t$ is the winning chain} \label{eq:min-reward}
\end{equation}
\end{definition}
\textbf{Discussion:}
\vspace*{-0.2cm}
\begin{itemize}
\item The $\delta$-SCR property guarantees that recruiter of each agent on the winning chain gets a certain fraction of the agent's reward. This property will encourage an agent to propagate the message to her acquaintances even though she may not execute the task by herself. This would result in rapid growth of the network which is desirable in many settings.
\item On the other hand, $\gamma$-SEC ensures that the reward to the winner remains larger than a fraction of the total reward. This works as a motivation for any agent to spend effort on executing the task by herself.
\end{itemize}
In what follows, we characterize the space of mechanisms satisfying these properties.
\section{Cost Critical Tasks} \label{sec:cost-critical}
In this section, we design crowdsourcing mechanisms for the atomic tasks where the planner's objective is to minimize total cost of executing the task.
%
\begin{definition}[\texttt{MINCOST} over $\mathscr{C}$]
 A reward mechanism $R$ is called \emph{\texttt{MINCOST}} over a class of mechanisms $\mathscr{C}$, if it minimizes the total reward distributed to the participants in the winning chain. That is, $R$ is \emph{\texttt{MINCOST}} over $\mathscr{C}$, if
 \begin{eqnarray}
 & R \in \argmin_{R' \in \mathscr{C}} \sum_{k=1}^t R'(k,t) , \quad \forall t. \label{eq:mincost}
 \end{eqnarray}
\end{definition}
We will show that the \texttt{MINCOST} mechanism over the space of $\epsilon$-DSP, $\delta$-SCR, and BB properties is completely characterized by a simple geometric mechanism, defined below.
\begin{definition}[$(\gamma,\delta)$-Geometric Mechanism, $(\gamma,\delta)$-GEOM]
\label{def:gam-delta-geom}
 This mechanism gives $\gamma$ fraction of the total reward to the winner and geometrically decreases the rewards towards root with the factor $\delta$. For all $t$, $R(t,t) = \gamma \cdot R_{\max}$;  $R(k,t) = \delta \cdot R(k+1,t) = \delta^{t-k} \cdot R(t,t)= \delta^{t-k} \cdot \gamma R_{\max}, \ k \leq t-1$.
\end{definition}
\subsection{Characterization Theorem for MINCOST}
Now, we will show that $(\gamma,\delta)$-Geometric mechanism characterizes the space of \texttt{MINCOST} mechanisms satisfying $\epsilon$-DSP, $\delta$-SCR, $\gamma$-SEC, and BB.
We start with an intermediate result.
\begin{lemma}
 \label{lem:scr-bb}
 A mechanism is $\delta$-SCR, $\gamma$-SEC and BB only if $\gamma \leq 1-\delta$.
\end{lemma}
\begin{prf}
 Suppose $\gamma > 1-\delta$. Then by $\delta$-SCR, we have,
 \begin{align}
  \mbox{$\sum_{k=1}^t R(k,t)$} &\geq (1+\delta + \dots + \delta^{t-1}) \cdot R(t,t) \label{eq:dscr}\\
  &\geq (1 + \delta + \dots + \delta^{t-1}) \cdot \gamma R_{\max} \label{eq:gsec}\\
  &> (1 + \delta + \dots + \delta^{t-1}) (1-\delta) R_{\max} \nonumber
 \end{align}
 This holds for all $t \geq 1$. It must hold for $t \to \infty$. Hence, $\lim_{t \to \infty} \sum_{k=1}^t R(k,t) > \frac{1}{1-\delta} \cdot (1-\delta) R_{\max} = R_{\max}.$ Which is a contradiction to BB.
\end{prf}
\begin{theorem}
\label{thm:main-full}
If $\delta \leq \min \{1-\gamma, \frac{\epsilon}{1+\epsilon}\}$, a mechanism is \emph{\texttt{MINCOST}} over the class of mechanisms satisfying $\epsilon$-DSP, $\delta$-SCR, $\gamma$-SEC, and BB iff it is $(\gamma,\delta)$-GEOM mechanism.
\end{theorem}
\begin{prf} ($\Leftarrow$) It is easy to see that $(\gamma,\delta)$-GEOM is $\delta$-SCR and $\gamma$-SEC by construction. It is also BB since $\delta \leq 1-\gamma$ or $\gamma \leq 1-\delta$. For the $\epsilon$-DSP property, we see that the following expression,
  \begin{align*}
  \lefteqn{\mbox{$\sum_{i=0}^n R(k+i,t+n) =  \sum_{i=0}^n \delta^{t+n-k-i} \cdot R(t+n,t+n)$}} \\
      &= \delta^{t-k} \cdot (1+\delta + \dots + \delta^n) \cdot \gamma R_{\max} \\
      &= R(k,t) \cdot (1+\delta + \dots + \delta^n) \\
      &\leq R(k,t) \cdot \frac{1}{1-\delta} \leq (1+\epsilon) R(k,t), \mbox{ as } \delta \leq \frac{\epsilon}{1+\epsilon}.
 \end{align*}
Also for a given $\delta$ and $\gamma$, this mechanism minimizes the total cost as it pays each node the minimum possible reward. Thus, $\delta$-GEOM mechanism is \texttt{MINCOST} over $\epsilon$-DSP, $\delta$-SCR, $\gamma$-SEC, and BB.

($\Rightarrow$) Since $\delta \leq 1-\gamma$, from Lemma~\ref{lem:scr-bb}, we see that $\delta$-SCR, $\gamma$-SEC, and BB are satisfiable. In addition the objective of the mechanism designer is to minimize the total reward ($R_{total}$) given to the winning chain.
\begin{align*}
 R_{total} &= \mbox{$\sum_{k=1}^t R(k,t)$} \stackrel{\text{Eq.~\ref{eq:gsec}}}{\geq} (1+\delta + \dots + \delta^{t-1}) \cdot \gamma R_{\max}
\end{align*}
We require a mechanism that is also $\epsilon$-DSP and minimizes the above quantity. Let us consider a mechanism $R_1$ that pays the leaf an amount of $\gamma R_{\max}$ and any other node at depth $k$, an amount $\delta^{t-k} \gamma R_{\max}$. We ask the question if this mechanism is $\epsilon$-DSP. This is because if this is true, then there cannot be any other mechanism that  minimizes the cost, as this achieves the lower bound of $R_{total}$. To check for $\epsilon$-DSP of this mechanism, we consider the following expression.
\begin{align*}
  \lefteqn{\mbox{$\sum_{i=0}^n R_1(k+i,t+n) =  \sum_{i=0}^n \delta^{t+n-k-i} \cdot R_1(t+n,t+n)$}} \\
      &= \delta^{t-k} \cdot (1+\delta + \dots + \delta^n) \cdot \gamma R_{\max} \\
      &= R_1(k,t) \cdot (1+\delta + \dots + \delta^n) \\
      &\leq R_1(k,t) \cdot \frac{1}{1-\delta} \leq (1+\epsilon) R_1(k,t) \quad \mbox{ since } \delta \leq \frac{\epsilon}{1+\epsilon}
 \end{align*}
implying $R_1$ is also $\epsilon$-DSP. Hence, $R_1$ is \emph{the} \texttt{MINCOST} mechanism over $\epsilon$-DSP, $\delta$-SCR, $\gamma$-SEC, and BB. Note, $R_1$ is precisely the $(\gamma,\delta)$-GEOM mechanism.
\end{prf}\\

\begin{figure}
\begin{minipage}{0.48\textwidth}
\centering
 \includegraphics[width = \textwidth]{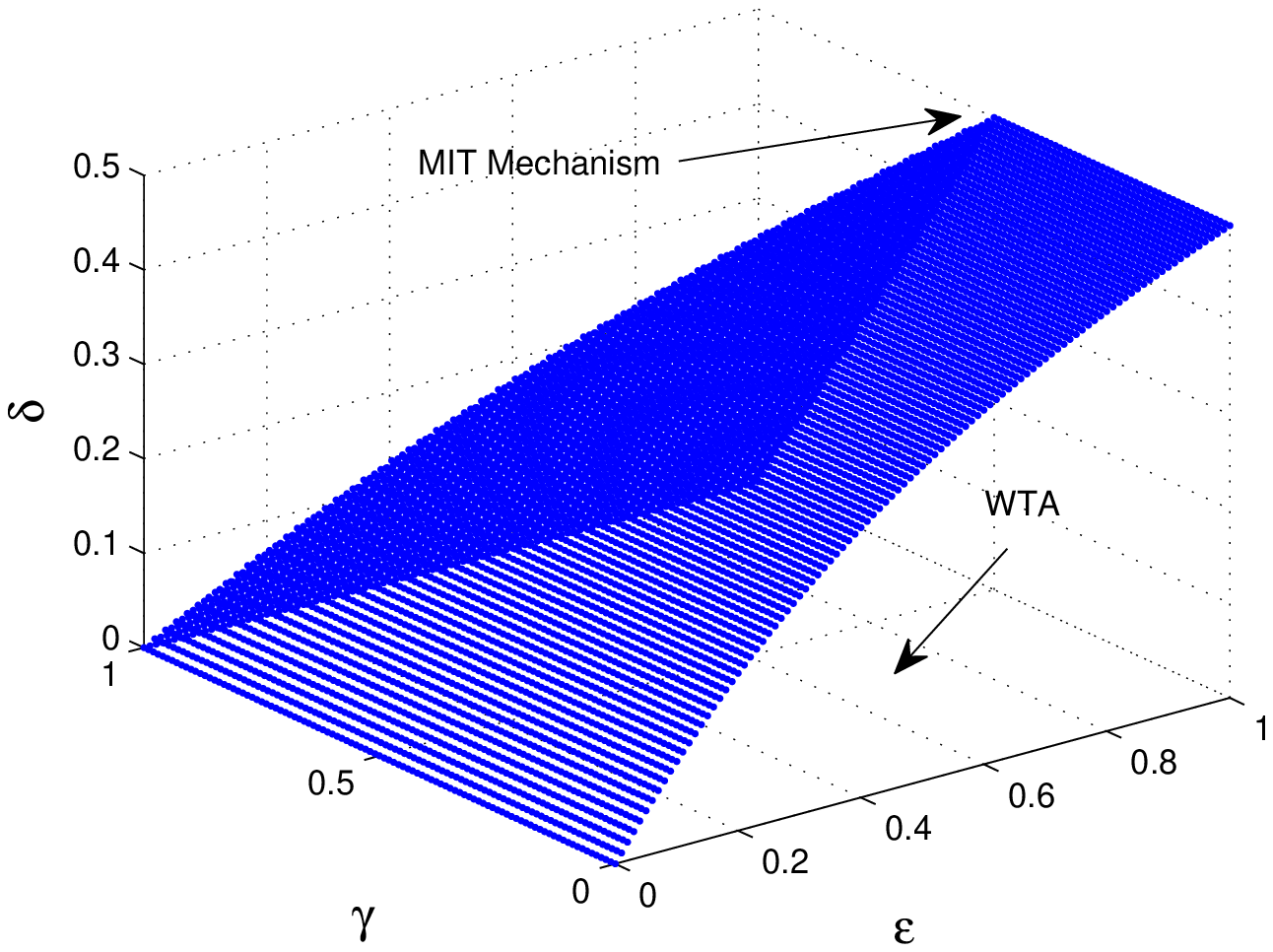}
 \caption{The space of $(\delta, \epsilon, \gamma)$ tuples characterized by Theorem~\ref{thm:main-full} holds}
 \label{fig:feasible-region}
\end{minipage}
\hspace*{0.1cm}
\begin{minipage}{0.48\textwidth}
\centering
    \includegraphics[width=0.9\textwidth]{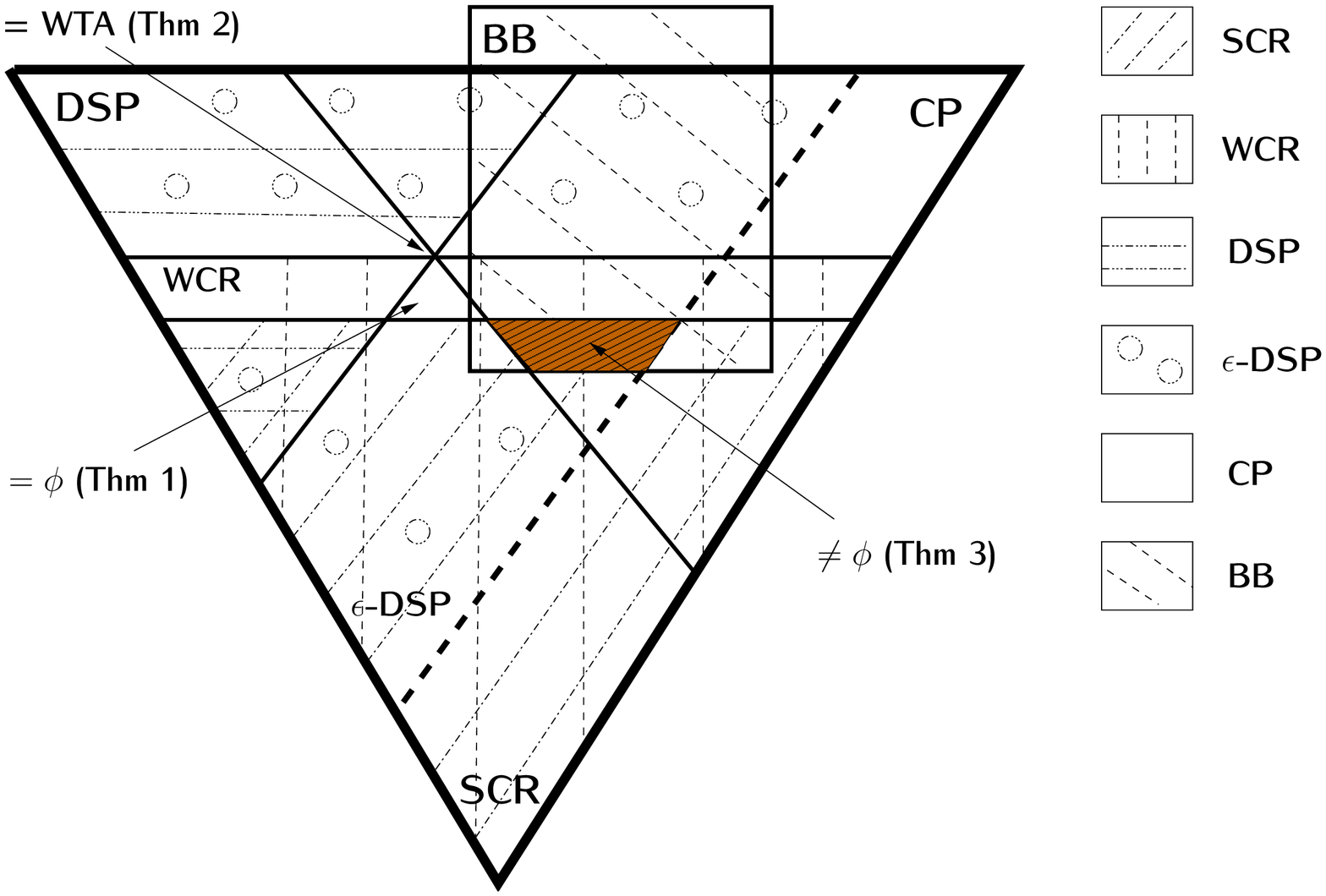}
    \caption{Graphical illustration of a part of the results in this paper.}
 \label{fig:space}
\end{minipage}
\end{figure}

\textbf{Discussions:}
\vspace*{-0.2cm}
\begin{itemize}
\item  Note, $(\gamma,\delta)$-GEOM mechanism additionally satisfies CP. The proof for this is given in Appendix A.
\item Theorem~\ref{thm:main-full} imposes a constraint on the values of the parameters $\delta$, $\epsilon$, and $\gamma$, for which the characterization result holds. Let us define,
\[\mathscr{E} = \{(\delta, \epsilon, \gamma) : \delta \leq \min \{1-\gamma, \frac{\epsilon}{1+\epsilon} \}\}.\]
The space $\mathscr{E}$ is graphically illustrated in the Figure~\ref{fig:feasible-region}.
\item The set $\mathscr{E}$ of $(\delta, \epsilon, \gamma)$ tuples that are characterizable via Theorem~\ref{thm:main-full} is the space below the shaded region in Figure~\ref{fig:feasible-region}. MIT mechanism ($\epsilon = 1, \delta = 0.5, \gamma = 0.5$) and the WTA mechanism ($\delta = 0$, the floor of the space in the figure above) are special cases. Theorem~\ref{thm:main-full} says that if $(\delta, \epsilon, \gamma) \in \mathscr{E}$, the characterization result in that theorem holds.
\end{itemize}
\section{Time Critical Tasks} \label{sec:time}
In applications where the faster growth of network is more important than maximizing the surplus, the designer can spend the whole budget in order to incentivize participants to either search for the answer or forward the information quickly among their acquaintances. In such settings, we can design mechanisms which aim to maximize reward of the leaf node of the winning chain. In this section, we show that such kind of mechanisms with the same fairness guarantees can also be characterized by a similar mechanism that exhausts the budget even for a finite length of the winning chain. In what follows, we define the design goal and a specific geometric mechanism.
\begin{definition}[\texttt{MAXLEAF} over $\mathscr{C}$]
 A reward mechanism $R$ is called \emph{\texttt{MAXLEAF}} over a class of mechanisms $\mathscr{C}$, if it maximizes the reward of the leaf node in the winning chain. That is, $R$ is \emph{\texttt{MAXLEAF}} over $\mathscr{C}$, if
 \begin{equation}
  R \in \argmax_{R' \in \mathscr{C}} R'(t,t) , \quad \forall t. \label{eq:maxleaf}
 \end{equation}
\end{definition}

\begin{definition}[$\delta$-Geometric mechanism, $\delta$-GEOM]
\label{def:delta-geom}
  This mechanism gives $\frac{1-\delta}{1-\delta^t}$ fraction of the total reward to the winner and geometrically decreases the rewards towards root with the factor $\delta$, where $t$ is the length of the winning chain. For all $t$, $R(t,t) = \frac{1-\delta}{1-\delta^t} \cdot R_{\max}$; $R(k,t) = \delta \cdot R(k+1,t) = \delta^{t-k} \cdot R(t,t), \ k \leq t-1$.
\end{definition}
\vspace{-2mm}
\subsection{Characterization Theorem for MAXLEAF}
\vspace{-2mm}
\begin{theorem}
\label{thm:maxleaf}
If $\delta \leq \frac{\epsilon}{1+\epsilon}$, a mechanism is \emph{\texttt{MAXLEAF}} over the class of mechanisms satisfying $\epsilon$-DSP, $\delta$-SCR, and BB iff it is $\delta$-GEOM mechanism.
\end{theorem}
\begin{prf}
 ($\Leftarrow$) By construction, the $\delta$-GEOM mechanism is $\delta$-SCR and BB for all $t$. It is also $\epsilon$-DSP, as,
\begin{align*}
  &\mbox{$\sum_{i=0}^n R(k+i,t+n) = \sum_{i=0}^n \delta^{t+n-k-i} \cdot R(t+n,t+n)$}  \\
      &= \delta^{t-k} \cdot (1+\delta + \dots + \delta^n) \cdot R(t+n,t+n) \\
      &= \delta^{t-k} R(t,t) \cdot \frac{R(t+n,t+n)}{R(t,t)} \cdot \frac{1-\delta^{n+1}}{1-\delta}\\
        &= R(k,t) \cdot \frac{R(t+n,t+n)}{R(t,t)} \cdot \frac{1-\delta^{n+1}}{1-\delta}\\
        &= R(k,t) \cdot \frac{\frac{1-\delta}{1-\delta^{t+n}} \cdot R_{\max}}{\frac{1-\delta}{1-\delta^t} \cdot R_{\max}} \cdot \frac{1-\delta^{n+1}}{1-\delta}
 \end{align*}
 \begin{align*}
      &= R(k,t) \cdot \frac{1-\delta^{n+1}}{1-\delta^{t+n}} \cdot \frac{1-\delta^t}{1-\delta}.
 \end{align*}
Since $\frac{1-\delta^{n+1}}{1-\delta^{t+n}} \uparrow n$ and $\frac{1-\delta^t}{1-\delta} \uparrow t$, we can take limits as $n \to \infty$ and $t \to \infty$ respectively to get an upper bound on the quantity of the RHS, which gives,
\begin{align*}
 \sum_{i=0}^n R(k+i,t+n) &= R(k,t) \cdot \frac{1}{1-\delta} \leq (1+\epsilon) \cdot R(k,t),
 \end{align*}
 since $\delta \leq \frac{\epsilon}{1+\epsilon}$. Hence this is $\epsilon$-DSP. Suppose this is not \texttt{MAXLEAF}. Then $\exists$ some other mechanism $R'$ in the same class that pays $R'(t,t) > \frac{1-\delta}{1-\delta^t} \cdot R_{\max}$. Since, $R'$ is also $\delta$-SCR,
\begin{align*}
  \lefteqn{\mbox{$\sum_{k=1}^t R'(k,t) \geq (1+\delta + \dots + \delta^{t-1}) \cdot R'(t,t)$}} \\
      &= \frac{1-\delta^t}{1-\delta} \cdot R'(t,t) > \frac{1-\delta^t}{1-\delta} \cdot \frac{1-\delta}{1-\delta^t} \cdot R_{\max} = R_{\max},
 \end{align*}
which is a contradiction to BB. Hence proved.

 ($\Rightarrow$) Let $R$ be a mechanism that is \texttt{MAXLEAF} over the class of mechanisms satisfying $\epsilon$-DSP, $\delta$-SCR, and BB. Hence,
 \begin{align}
  R_{\max} &\geq \sum_{k=1}^t R(k,t) \stackrel{\text{Eq.~\ref{eq:dscr}}}{\geq} \frac{1-\delta^t}{1-\delta} \cdot R(t,t) \nonumber \\
 &\Rightarrow \quad R(t,t) \leq \frac{1-\delta}{1-\delta^t} \cdot R_{\max}, \quad \mbox{for all $t$.} \label{eq:nec-dscr-bb}
 \end{align}
 The first and second inequalities arise from BB and $\delta$-SCR respectively. Now, from the $\epsilon$-DSP condition of $R$, we get, for all $n, t, k \leq t$,
\begin{align*}
 (1+\epsilon) R(k,t) &\geq \mbox{$\sum_{i=0}^n R(k+i,t+n)$} \\
		    &\geq \mbox{$\sum_{i=0}^n \delta^{t+n-k-i} \cdot R(t+n,t+n)$} \\
		    &= \delta^{t-k} \cdot (1+\delta + \dots + \delta^n) \cdot R(t+n,t+n),
\end{align*}
where the second inequality comes from $\delta$-SCR of $R$. Rearranging, we obtain,
\begin{equation}
 1+\epsilon \geq \delta^{t-k} \cdot \frac{1-\delta^{n+1}}{1-\delta} \cdot \frac{R(t+n,t+n)}{R(k,t)} \label{eq:nec-edsp}
\end{equation}
Since this is a necessary condition for any $k \leq t$, it should hold for $k = t$ in particular. Using this in Equation~\ref{eq:nec-edsp} the necessary condition becomes,
\begin{equation}
 1+\epsilon \geq \frac{1-\delta^{n+1}}{1-\delta} \cdot \frac{R(t+n,t+n)}{R(t,t)} \label{eq:nec-edsp-1}
\end{equation}
Now, we have two conditions on $R(t+n,t+n)$ as follows.
\begin{align}
 \lefteqn{R(t+n,t+n) \leq (1+\epsilon) \cdot \frac{1-\delta}{1-\delta^{n+1}} \cdot R(t,t)} \nonumber \\
	  &\stackrel{\text{Eq.~\ref{eq:nec-dscr-bb}}}{\leq} \underbrace{(1+\epsilon) \cdot \frac{1-\delta}{1-\delta^{n+1}} \cdot \frac{1-\delta}{1-\delta^t} \cdot R_{\max}}_{=: A(n,t)} \label{eq:cond-1}
\end{align}
and using Eq.~\ref{eq:nec-dscr-bb} directly on $R(t+n,t+n)$, we get,
\begin{equation}
 R(t+n,t+n) \leq \underbrace{\frac{1-\delta}{1-\delta^{t+n}} \cdot R_{\max}}_{=: B(n,t)} \label{eq:cond-2}
\end{equation}
It is clear that to satisfy $\delta$-SCR, $\epsilon$-DSP and BB, it is necessary for $R$ to satisfy, $$R(t+n,t+n) \leq \min_{n,t} \{A(n,t), B(n,t)\}.$$
We can show the following bounds for the quantity $\frac{B(n,t)}{A(n,t)}$, which we skip due to space constraints.
\begin{equation}
 \frac{1}{1+\epsilon} \leq \frac{B(n,t)}{A(n,t)} \leq \frac{1}{(1+\epsilon)(1-\delta)}. \label{eq:ratio-bound}
\end{equation}
Since $\delta \leq \frac{\epsilon}{1+\epsilon}$, we see that the upper bound $\frac{1}{(1+\epsilon)(1-\delta)} \leq 1$. Hence, $A(n,t)$ uniformly dominates $B(n,t)$, $\forall \ n,t$. Hence, $R(t+n,t+n) \leq B(n,t)$. Since $R$ is also \texttt{MAXLEAF}, equality must hold and it must be true that,
\begin{equation}
 R(t,t) = \frac{1-\delta}{1-\delta^t} \cdot R_{\max}, \forall \ t. \label{eq:maxleaf-1}
\end{equation}
Also, since $R$ is BB, it is necessary that,
\begin{equation}
 R(k,t) = \delta^{t-k} \cdot R(t,t), \quad k \leq t-1. \label{eq:maxleaf-2}
\end{equation}
This shows that $R$ has to be $\delta$-GEOM.
\end{prf}\\
\textbf{Discussion:} A $\delta$-GEOM mechanism also satisfies CP property. The proof for this is given in Appendix B.
\section{Conclusions and Future Work} \label{sec:concl}
In this paper, we have studied the problem of designing manipulation free crowdsourcing mechanisms for atomic tasks under the cost critical and time critical scenarios. We have motivated the need for having CP as an additional property of the mechanism beyond what already exists in the literature. Starting with an impossibility result, we have developed mechanisms for both cost and time critical scenarios which satisfy CP property along with weaker versions of other desirable properties (all under dominant strategy equilibrium).
Figure~\ref{fig:space} summarizes part of the results presented in this paper. The three corners of the triangular space are used to denote the space of mechanisms satisfying properties DSP, CP, and SCR, respectively. Space satisfying $\epsilon$-DSP is a super set of the DSP space and is shown dotted in the figure. The figure shows that no mechanism can satisfy SCR, DSP, and CP, and the only mechanism that satisfies WCR, DSP, and CP is the Winner Takes All (WTA, defined formally in the paper) mechanism. Once we relax DSP to $\epsilon$-DSP, it is possible to satisfy it along with SCR, CP, and Budget Balance.
We find that there is a scope for further investigation in the cost-critical setting, but for the time-critical scenario, our results are tight and characterize the entire space of mechanisms. We would characterize the  complementary scenarios of our results in the cost-critical setting in our future work.

\bibliographystyle{plainnat}
\bibliography{/home/swaprava/Documents/PhD-Work/Research/master16072012}
\newpage
\appendix
\renewcommand\thesection{Appendix \Alph{section}:} 
\section{Proof for $(\gamma, \delta)$-GEOM Mechanism is Collapse-Proof}
 For the mechanism $(\gamma, \delta)$-GEOM, the reward of the leaf node is independent of the length of the winning chain, i.e., $t$. Hence, we can use the same proof technique used in Theorem~\ref{thm:existence} to show that $(\gamma, \delta)$-GEOM is CP. The steps are as follows.
  \begin{align*}
  \lefteqn{\mbox{$\sum_{i=0}^p R(k+i,t) = \sum_{i=0}^p \delta^{t-k-i} \cdot R(t,t)$}} \\
      &= \mbox{$\sum_{i=0}^{p-1} \delta^{t-k-i} \cdot R(t,t)$} + \underbrace{\delta^{t-k-p} \cdot R(t,t)}_{ = R(k,t-p)} \geq R(k,t-p)
 \end{align*}
 This shows that this mechanism is CP.
\section{Proof for $\delta$-GEOM Mechanism is Collapse-Proof}
 In order to prove the claim, we need to show that,
 \begin{align}
  & &\sum_{i=0}^{p} R(k+i, t) &\geq R(k,t-p) \quad \forall k +p \leq t, \forall t  \nonumber \\
  &\Rightarrow& \sum_{i=0}^{p} \delta^{t-k-i} \cdot R(t,t) &\geq \delta^{t-k-p} \cdot R(t-p,t-p) \nonumber \\
  &\Rightarrow& \sum_{i=0}^{p} \delta^{t-k-i} \cdot \frac{1-\delta}{1-\delta^t} \cdot R_{\max} &\geq \delta^{t-k-p} \cdot \frac{1-\delta}{1-\delta^{t-p}} \cdot R_{\max} \nonumber \\
  &\Rightarrow& (1-\delta^{t-p}) \sum_{i=0}^{p} \delta^{t-k-i} &\geq \delta^{t-k-p} (1-\delta^t)  \nonumber \\
  &\Rightarrow& (1-\delta^{t-p}) (\delta^{t-k-p} + \dots + \delta^{t-k}) &\geq \delta^{t-k-p} - \delta^{2t-k-p}  \nonumber \\
  &\Rightarrow& \cancel{\delta^{t-k-p}} + \dots + \delta^{t-k} - (\delta^{2t-2p-k} + \dots & \nonumber \\
  &&     + \cancel{\delta^{2t-p-k}}) &\geq \cancel{\delta^{t-k-p}} - \cancel{\delta^{2t-k-p}}  \nonumber \\
  &\Rightarrow& (\delta^{t-k} + \dots + \delta^{t-k-p+1}) - (\delta^{2t-2p-k} + \dots & \nonumber \\
  && 	\delta^{2t-p-k-1}) &\geq 0. \label{eq:ascending}
 \end{align}
Where both the terms within the parentheses on the LHS are in ascending order. We can rewrite inequality (\ref{eq:ascending}) as,
 \begin{align*}
  && \delta^{t-k-p} \left[(\delta^p - \delta^{t-p})+ (\delta^{p-1} - \delta^{t-p+1}) + \dots + (\delta^{1}- \delta^{t-1}) \right] &\geq 0.
 \end{align*}
Let us define, $a^t_k := \delta^k - \delta^{t-k}$. Therefore, from the above inequality, we need to show that,
 \begin{align}
  \sum_{k=1}^p a^t_k &\geq 0. \label{eq:more-ascending}
 \end{align}
This is a partial sum of the complete sum, $\sum_{k=1}^{t-1} a^t_k$ which equals zero. Since, $\delta < 1$, we also see that,
 \begin{align*}
  a^t_k &\geq 0 & \forall & k = 1, \dots, \lfloor t/2 \rfloor, \\
  a^t_k &\leq 0 & \forall & k = \lfloor t/2 \rfloor + 1, \dots, t-1.
 \end{align*}
For any $p$, the partial sum would be non-negative. So, inequality (\ref{eq:more-ascending}) holds for any $p$. Therefore, we have shown that $\delta$-GEOM is CP.
\end{document}